# Resolving Conductor-to-Plasma Transition Through Low-Energy Electrical Explosion of metallic foils

Aditya Nandan Savita[1,2], Sambaran Pahari[1,2], Neeraj Shiv [1], IVV Suryaprasad[1]

[1]Computational Analysis Division, Bhabha Atomic Research Centre, Visakhapatnam, India-531011

[2]Homi Bhabha National Institute, Anushaktinagar, Mumbai, Maharashtra, India- 400094

*adityans@barc.gov.in*

**Abstract:**

Electrical explosion of metallic foils involves a rapid transition from solid-state conduction through melting and vaporization to formation of an ionized plasma. Resolving intermediate processes is challenging because these transformations occur over strongly overlapping timescales. In this work, to examine temporal evolution of conductor-to-plasma transition, electrical explosion of Al foil is investigated under two controlled energy-deposition regimes. At higher energy deposition, synchronized current, voltage, and fast-framing optical measurements capture explosive destruction of foil and subsequent plasma formation. However, underlying intermediate processes remain temporally compressed. To resolve these processes, experiments are performed at a lower energy-deposition rate while maintaining comparable circuit timescales. Under these conditions, characteristic features in the electrical waveforms allow successive stages of heating, melting, liquid-phase evolution, vaporization, and vapor discharge to be delineated. Experimentally determined action integrals associated with observed transitions are compared with predictions based on the Tucker–Toth specific-action model and available literature. A particularly notable observation is a sharp voltage turnaround immediately preceding current restrike, occurring simultaneously with increasing current and reproducibly associated with plasma formation. Although its physical origin remains unresolved, the feature suggests a rapid nonlinear reorganization during final transition from exploding conductor and metal vapor to plasma conduction. Results demonstrate that controlled energy-deposition rate provides an effective means of temporally resolving processes otherwise compressed and offers a useful approach for investigating the dynamics preceding plasma formation in electrically exploded foils.

## 1. Introduction:

Electrical explosion of metallic conductors has been investigated extensively for several decades (1) (2) (3) (4) (5) (6) (7) (8) (9) (10) (11) because of its relevance to fundamental studies and applications related to rapidly heated matter, pulsed-power technology, shock generation, and plasma formation. In these experiments, a large electrical current is passed through a metallic conductor over a short timescale, resulting in intense Joule heating and the successive transformation of the material from a solid conductor to liquid, vapor, and ultimately an ionized plasma. Electrical explosions are most commonly studied using two conductor geometries: cylindrical wires (12) (13) (10) (14) (15), typically ranging from a few tens to several hundred micrometres in diameter, and thin metallic foils (14) (11) (17). Although both geometries undergo the same fundamental sequence of rapid heating, phase transformation, vaporization, and plasma formation, their explosion dynamics, degree of heating uniformity, and applications can differ significantly.

From a fundamental perspective, electrical explosion provides a laboratory system in which the unique nonequilibrium pathway from condensed-matter transport to plasma transport can be followed dynamically through coupled electrical and optical observables. During the discharge, the mechanism of charge transport evolves continuously from metallic conduction through liquid and vapor states to conduction in an ionized medium, while the density, temperature, geometry, and electrical boundary conditions change simultaneously. Resolving this evolution is therefore relevant to the physics of rapid phase transformation, nonequilibrium transport, metal-vapor ionization, plasma formation, and nonlinear coupling between matter and an external circuit. The observed pre-restrike dynamics are particularly significant because they may represent a distinct transition between the loss of metallic conduction and the establishment of a self-sustaining plasma current channel. Resolving this interval experimentally provides a basis for connecting material-phase evolution with the physics of plasma formation in a strongly driven, transient system.

Beyond its fundamental significance, resolving the temporal sequence of phase transformations during electrical explosion is relevant to a broad range of pulsed-power and plasma technologies (3) (15) (16) (17) (18) (19). The performance of exploding-conductor opening switches (19) (16), flyer-plate and shock-generation systems (6) (9), pulsed plasma sources, warm-dense-matter experiments (20) (21), and materials-processing techniques depends not only on the total deposited energy but also on the temporal pathway through melting, vaporization, loss of metallic continuity, and plasma formation (22). The identification of reproducible electrical markers associated with these transformations may therefore provide useful constraints for Multiphysics modelling, improved control of energy deposition, and the development of diagnostic or feedback strategies for pulsed-power-driven material and plasma systems.

Despite the extensive literature on electrical explosions, the detailed temporal evolution connecting the initially solid conductor to the final plasma state remains difficult to resolve experimentally. Prior to complete electrical explosion, the material successively undergoes heating of the solid phase, melting, heating of the liquid phase, and vaporization. Following the loss of metallic continuity, electrical breakdown of the generated metal vapor can lead to arc formation, ionization, and eventually the establishment of a conducting plasma. In practice, these processes occur rapidly and overlap to varying degrees, making their individual electrical signatures difficult to identify.

In the present work, the electrical explosion of Al foils is investigated under deliberately controlled energy-deposition conditions to temporally resolve these successive processes. The experiments are performed over relatively long discharge timescales, of the order of hundreds of microseconds, compared with conventional rapidly driven electrical explosions. Two regimes of energy deposition are examined. At higher input energy, synchronized electrical and optical measurements are used to follow the explosive destruction of the foil and the subsequent formation and evolution of a luminous plasma. Under these conditions, the processes preceding plasma formation are compressed into a short interval and appear predominantly as a strong voltage excursion immediately preceding current restrike.

To investigate the physical processes contained within this rapid transition, a second set of experiments is performed at a lower energy-deposition rate while maintaining comparable circuit inductance, capacitance, and overall discharge timescales. The reduced rate of energy deposition spreads the explosion dynamics over a longer interval, enabling characteristic electrical signatures associated with successive phase transformations and vapor discharge to be identified. The experimentally derived phase-transition signatures are further examined using a simplified phenomenological description based on the specific-action approach of the Tucker–Toth model and compared with available literature data (23). Particular attention is given to the electrical dynamics immediately before restrike, which emerge as a potentially important marker of the transition from an exploding conductor and metal vapor to a conducting plasma.

The remainder of the paper is organized as follows. Section 2 describes the experimental setup and diagnostics. Section 3 presents the electrical and optical evolution during high-energy electrical explosion. Section 4 examines the temporally resolved phase-transformation signatures observed under lower-energy deposition conditions and compares them with a phenomenological model based on specific action. Section 5 summarizes the principal findings and discusses the implications of the observed pre-restrike dynamics for understanding conductor-to-plasma transitions during electrical explosions.

## 2. Experimental Setup and Diagnostics:

A 13µm thick, 0.04g of Al-foil was electrically exploded under atmospheric pressure to study the explosion dynamics. Experimental setup (schematic shown in Fig.1) consists of a pulsed power supply (12kJ, 960µF, 5kV) driven by energy storage capacitors. The electrical circuit is predominantly an LC circuit, with a total capacitance of 960 µF, inductance (of the foil and circuit) ~2.4µH and a negligible stray capacitance. The capacitor bank discharges into the foil in approximately 100 µs followed by a voltage (and current) reversal. The large capacitance ensures energy deposition on longer time scales ensuring uniform heating of Al-foil. The Al-foil was mounted between two parallel busbars and ends of the foil were held between copper surfaces to minimize contact resistances. Current Transformers (200kA, 1.5MHz) and Resistive Voltage Dividers (20kV, 70MHz) were used to measure time resolved current and voltage across the foil. A high-speed camera (at 1 lakh to 3.6 lakh frames per second) has been used to register foil radiation. In order to study the explosion dynamics and accompanying phase transitions the high-speed camera has been used in synchronization with other electrical diagnostics.

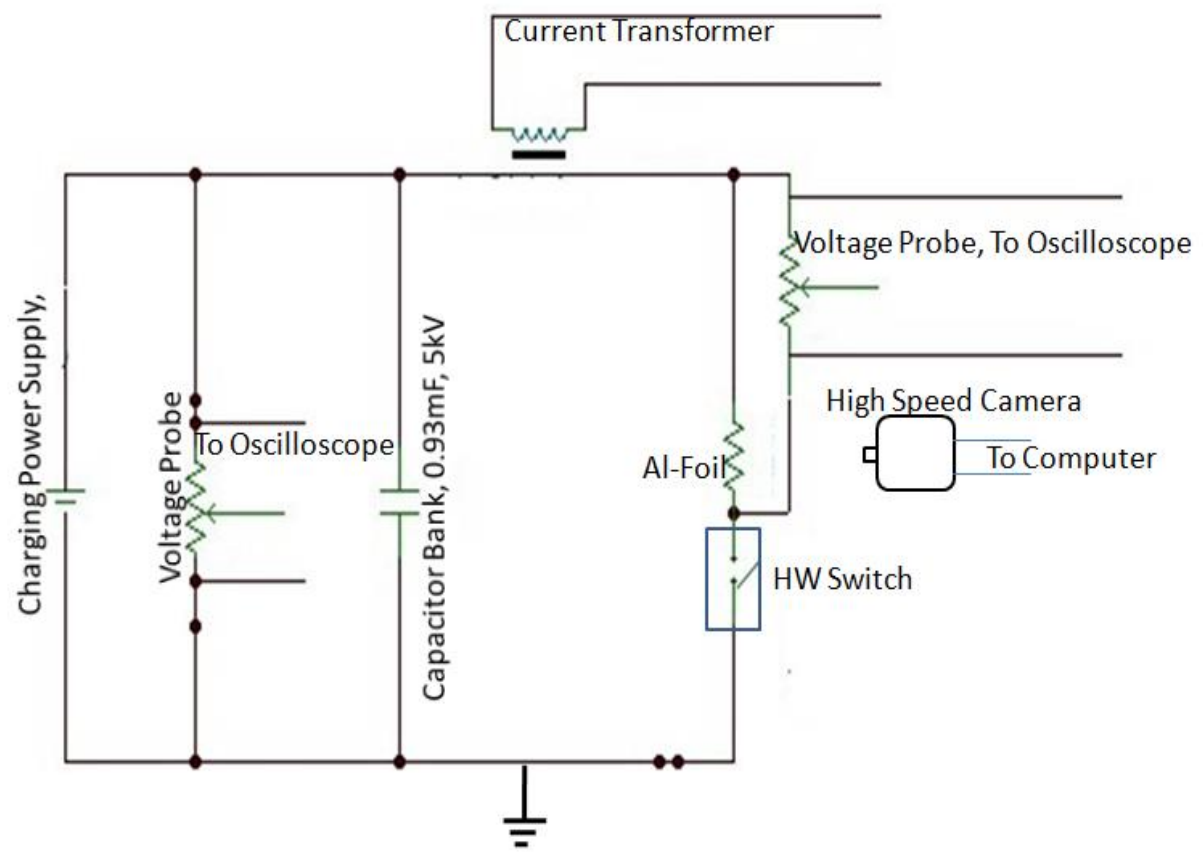

Figure 1.Experimental setup for electrical explosion of conductors

## 3. Electrical and Optical Evolution During High-Energy Electrical Explosion

A 13µm thick, 0.042g Al-foil was electrically exploded to form the plasma. The explosion was carried out by discharging the capacitor bank at ~1.9 kJ. At these energies, the electrical explosion of the Al foil represents the complete evolution of the conductor through several phase transitions into a plasma over a few tens of microseconds. The entire phenomena are recorded over 400 µs.

A typical discharge characteristic obtained in our experiments is shown in Fig 2. At 2 kV, maximum current of 23kA and peak voltage of 4.3kV is registered across the foil. Images of the foil radiation were captured using fast framing camera. The simultaneously recorded electrical waveforms and time-resolved framing images provide a comprehensive picture of this transformation. Electrical waveforms have been correlated with time resolved images in Fig2 and summarized in table 1.

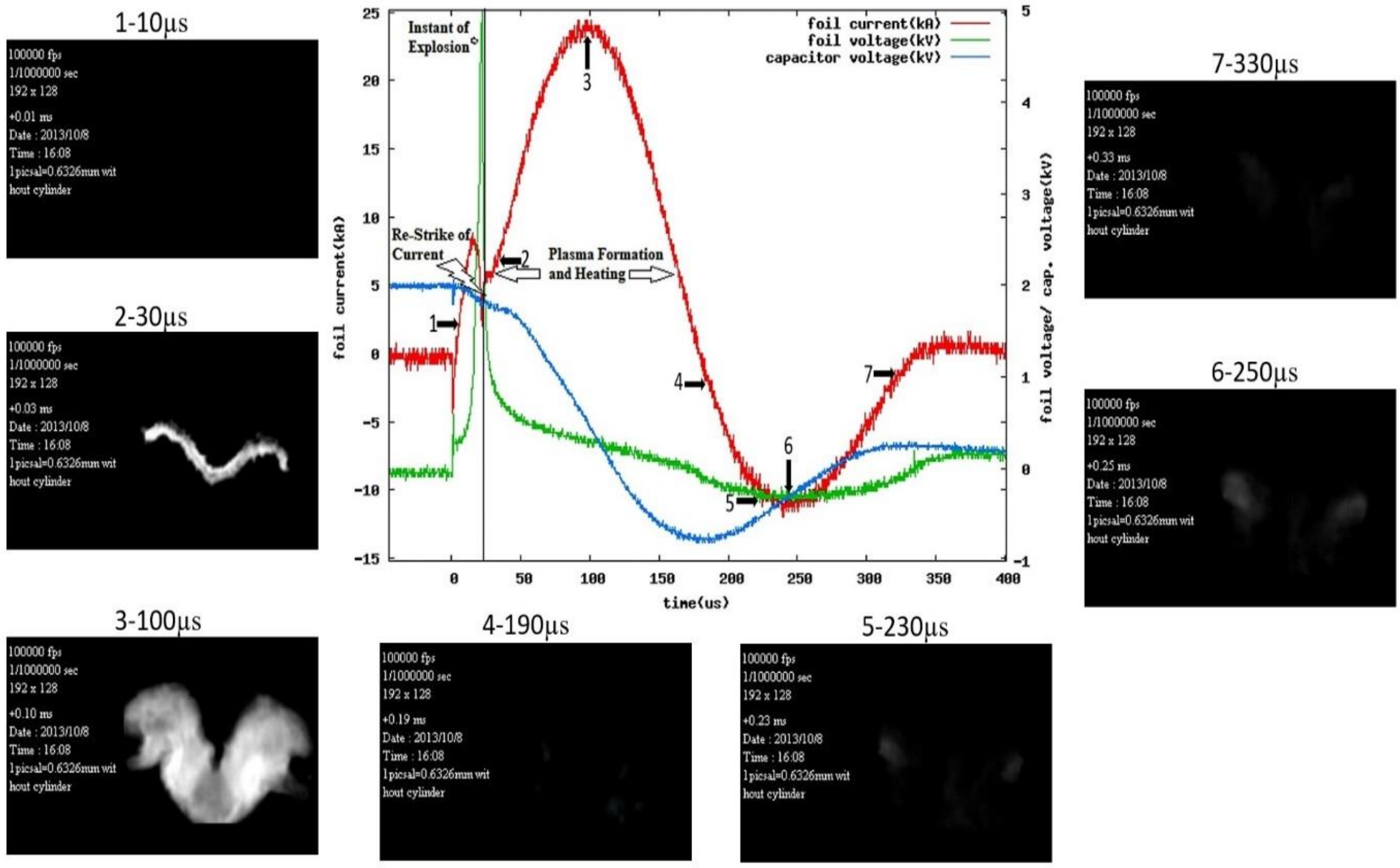

Figure 2.Correlation of foil radiation with electrical current flowing through the circuit

Table 1. Summary of correlation between electrical and optical evolutions during electrical explosion of Al-foil

| Image Number | Time (μs) | Observation in foil radiation | Observation by correlating with current and voltage signatures |
|---|---|---|---|
| 1 | 10 | No radiation from foil | Current starts flowing in the foil and joules heating of foil takes place |
| 2 | 30 | Intense radiation along the length of foil. Exploded foil has ionized but has not expanded significantly. | Onset-re-strike of current. Indicated Initiation of plasma formation |
| 3 | 100 | Foil radiation is maximum and has expanded in larger volume. | Peak of re-strike current. All energy of capacitors is now exhausted. |
| 4 | 190 | Foil radiation is diminished and only a few spots can be seen. | At this instant, current crosses to zero and capacitor is charged in reverse direction. Energy stored in plasma medium charges the capacitor back. |
| 5 | 230 | Foil radiation starts reappearing | Capacitor's energy re-ionizes the medium and forms plasma again. |
| 6 | 250 | Foil radiation reappearance peaks | Current in foil peaks in negative direction and capacitor energy is again zero. |
| 7 | 330 | Foil radiation diminished completely | Current approaches to zero and no-enough energy is left in capacitor to re-ionize the medium present between electrodes. |

Initially, the foil behaves as a low-resistance metallic conductor, resulting in a rapid rise of current with only a modest voltage appearing across the load. As Joule heating proceeds, energy is deposited at a sufficiently high rate that the foil rapidly traverses the solid, liquid and vapor phases. These individual phase transitions are not distinguishable electrically because the characteristic thermal processes are compressed into a very narrow temporal interval.

The most significant electrical feature of the discharge is the large voltage spike immediately preceding current restrike. This voltage excursion coincides with the disappearance of the metallic conductor and therefore represents the electrical signature of the explosive destruction of the foil. At this instant, the continuous metallic current path is interrupted by rapid vaporization and fragmentation of the foil, causing the load impedance to increase abruptly. Since the circuit inductance prevents an instantaneous change in current, a correspondingly large voltage develops across the exploding foil.

The framing images show that immediately following this voltage spike, intense optical emission appears almost simultaneously along the entire foil length while hydrodynamic expansion remains relatively limited. This observation indicates that the large electric field developed during the explosion initiates electrical breakdown of the freshly generated dense aluminium vapor. The discharge therefore undergoes a rapid transition from metallic conduction to plasma conduction, resulting in the observed current restrike. The voltage spike thus marks the termination of conduction through the condensed metal and the initiation of conduction through an ionized plasma channel.

Following re-strike, the discharge dynamics are governed primarily by the evolving plasma rather than the original metallic conductor. The current continues to increase while the plasma expands and its optical emission reaches maximum intensity, indicating efficient transfer of the remaining capacitor energy into plasma heating, ionization and expansion. At later times, the current decreases and the capacitor becomes reverse charged through energy returned from the plasma. The subsequent increase in reverse current is accompanied by renewed optical emission, suggesting re-ionization of the remaining metal vapor by the reversed capacitor voltage before the plasma eventually recombines and the discharge terminates.

These observations demonstrate that, under high-energy conditions, the electrical explosion is dominated by plasma dynamics immediately following foil destruction. However, the electrical waveforms provide only a single dominant voltage spike preceding restrike. Although it clearly signifies the transition from metallic conduction to plasma conduction, the underlying sequence of physical events remains unresolved. The discharge evidently undergoes a rapid reorganization from a condensed metallic conductor to an ionized plasma, but the individual contributions of melting, vaporization, metal-vapor formation, electrical breakdown and plasma channel establishment appear as a single composite electrical event as the numerous intermediate physical processes are compressed into an interval too short to be resolved.

This observation naturally motivates the use of a lower rate of energy deposition. By reducing the deposited energy while maintaining otherwise similar experimental conditions, the conductor-to-plasma transition can be temporally stretched, allowing the intermediate phase transformations to become electrically distinguishable. As will be shown in the following section, the lower-energy explosions effectively provide a time-resolved view of the same fundamental transition, enabling the successive stages of melting, vaporization, plasma initiation and current restrike to be identified individually. In this sense, the lower-energy experiments may be regarded as a "slow-motion" realization of the high-energy explosion, providing insight into the physical processes hidden within the apparently instantaneous voltage spike observed under high-energy conditions.

## 4. Electrical Signatures of Phase Transformation through Low-Energy Foil Explosion.

To investigate the physical processes hidden within the rapid conductor-to-plasma transition observed in the high-energy experiments, a second series of experiments was performed under conditions of reduced energy deposition. In these experiments, a similar 0.042 g Al foil was exploded using an input energy of approximately 0.57 kJ. The external circuit inductance and capacitance were maintained unchanged so that the overall discharge duration remained comparable to that of the high-energy experiments. Consequently, the energy delivered to the foil, which was reduced to approximately 30%, resulted in a correspondingly lower rate of energy deposition over essentially the same discharge timescale. The lower-energy experiments therefore provide a time-resolved view of the conductor-to-plasma transition, enabling the individual phase transformations occurring within the first ~100 μs of the electrical explosion to be identified and correlated with the measured current and voltage waveforms (Fig. 4).

**4.1 Phase changes observed from current-voltage waveform.** The discharge has been divided into 7 distinct regions and the trends in current, voltage (across the foil), resistance has been summarised in Table 2.

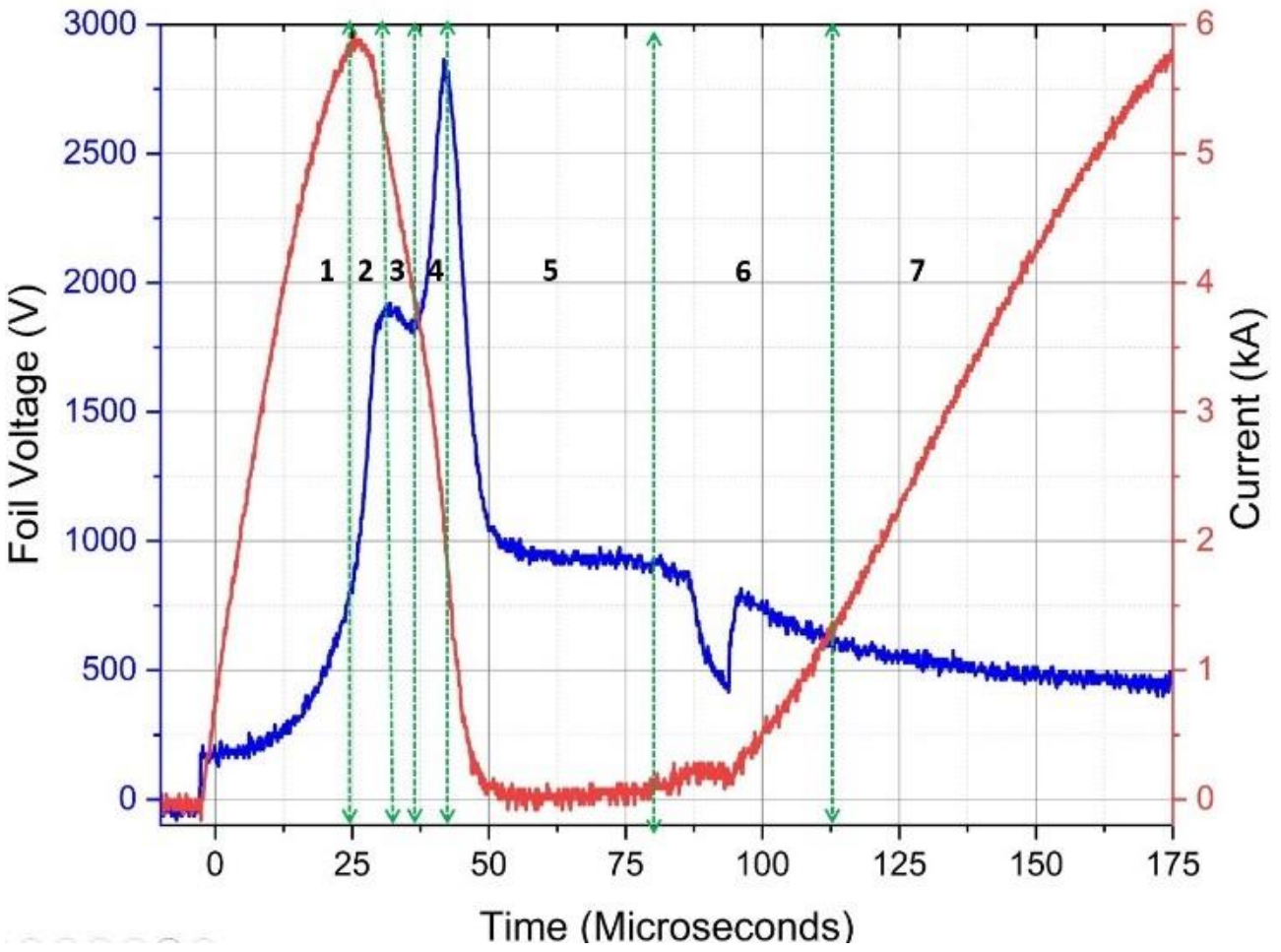

Figure 3.Electrical current and voltage across the foil at 0.57kJ capacitor bank energy

Table 2. Summary of voltage, current and resistance across the foil trends and drawn inference

| Region | current | Voltage across the foil | foil Res. | Dominating Process |
|---|---|---|---|---|
| 1 | **increases↑** | **increases↑** | ↑**increases** Gradual | Heating of Solid Foil |
| 2 | **decreases↓** | ↑**increases** Rapidly | ↑**increases** Rapidly | Phase Change: Melting of Solid Foil |
| 3 | **decreases↓** | **decreases↓** | **↑increases** Gradual | Heating of melted foil |
| 4 | **decreases↓** | **↑increases** Rapidly | **↑increases** Rapidly | Phase Change: Vaporization of hot melt foil |
| 5 | **decreases↓** | **decreases↓** | **increases↑** | Heating of vaporized foil/Arc formation |
| 6 | | Sharp dip | | Arc growth |
| 7 | **Current restrike** | | | Ionization |

The Fig. 3 and Table 2 together indicate the temporal evolution of the electrically exploded Al foil from a solid metallic conductor to a partially ionized vapor/plasma. The interpretation below is necessarily simplified because the different thermodynamic and electrical processes overlap in time and space. Nevertheless, under carefully chosen operating conditions (in this case, explosion of a 0.04 g Al foil with an input energy of about 0.57 J over hundreds of microseconds) the different stages become sufficiently separated that characteristic signatures can be reproducibly identified in the current and voltage waveforms.

As described in the previous section, initially the Al foil is in the solid metallic state and behaves as a low resistance conductor. As current begins to flow, Joule heating raises the foil temperature. During this stage (Region 1), both the circuit current and voltage increase gradually. The increase in foil temperature causes an increase in electrical resistivity of solid aluminium, leading to a gradual rise in foil resistance. Since the foil still remains electrically continuous, the current continues to rise with the applied circuit drive. This region therefore corresponds primarily to heating of the solid foil.

As the foil temperature approaches the melting point, the waveform enters Region 2. Here, the current decreases while the voltage across the foil rises rapidly. This behaviour is associated with the solid-to-liquid phase transition. Melting introduces structural disorder and rapidly increases resistivity. Consequently, the foil resistance rises sharply, producing a larger voltage drop across the foil while simultaneously suppressing current flow. Although melting is represented here as a separate stage, in practice melting occurs non-uniformly and overlaps with continued heating.

Once most of the foil has melted, the material enters a liquid-metal heating stage (Region 3). In this region both current and voltage decrease gradually. The molten aluminium continues to absorb energy and its temperature rises further toward the boiling point. The resistance continues to increase, though more gradually compared to the abrupt change during melting. Since the exploding foil system is strongly dynamic and the energy source begins transferring energy into an increasingly resistive load, the overall current decreases. The voltage also decreases because the rate of resistance growth is no longer sufficiently abrupt to sustain the earlier voltage rise.

Region 4 corresponds to vaporization of the hot liquid aluminium. Similar to the melting stage, vaporization produces a rapid structural and electrical transition. The current decreases sharply while the voltage rises rapidly. Formation of vapor cavities, fragmentation of the liquid column, and loss of metallic continuity drastically increase resistance. The electrical path becomes unstable and highly non-uniform. This rapid increase in resistance manifests as a strong voltage spike across the foil.

After substantial vaporization, the system enters Region 5, where heating of the metal vapor and arc formation dominate. The vaporized material is initially weakly conducting, but continued energy deposition heats and partially ionizes the vapor. Both current and voltage now decrease together while resistance continues to rise. In this stage, conduction no longer occurs through a continuous metal but through hot dense vapor and localized discharge channels. Arc precursors begin forming in regions where temperature and ionization become sufficiently high.

Region 6 represents arc growth. The initially localized conductive channels expand spatially and become increasingly conductive due to progressive ionization. The discharge transitions from a metal-vapor conduction process toward a plasma arc.

Finally, in region 7, a current re-strike is observed. This is the clearest signature of plasma formation and ionization. Once the vapor becomes sufficiently ionized, its conductivity increases dramatically, causing the circuit current to rise again. The plasma channel effectively provides a new low-resistance conduction path, resulting in restrike of current.

Two particularly important observations emerge from the waveforms:

1. During heating within a single phase, current and voltage vary in the same direction. During solid heating, both current and voltage increase because the foil remains electrically continuous while resistance increases gradually with temperature. During heating of the liquid or vapor phases, both current and voltage decrease because the circuit energy is increasingly dissipated in an expanding and dynamically changing medium, and the resistance evolution is slower and more distributed.

2. During phase transitions, current decreases while voltage increases. This is the characteristic signature of rapid resistance growth associated with phase change. Both melting and vaporization involve abrupt changes in material structure and conductivity. The sudden increase in resistance produces a larger voltage drop across the foil while simultaneously limiting current flow. Thus, opposite trends in current and voltage serve as indicators of major phase transitions in the exploding foil.

It must be emphasized that the above interpretation represents an idealized sequential description. In reality, the foil experiences strong spatial non-uniformities, localized overheating, hydrodynamic expansion, and coexistence of multiple phases. Melting, vaporization, and ionization can occur simultaneously in different regions of the foil. The reason these signatures become distinguishable under the present experimental conditions is that the deposited energy and timescale are sufficiently moderate to reduce excessive overlap between processes. When much larger energies are deposited over shorter timescales, the transitions occur too rapidly and simultaneously, making individual signatures difficult to resolve in the electrical waveforms.

### 4.2 Comparison with Tucker Toth Model

The current waveform alone can sometimes be ambiguous because current and voltage evolve simultaneously under circuit effects. Physically the specific action integral, usually written as $g=\int J^2\,dt$, (where J is the current density), is often evaluated to convert a complicated transient waveform into a thermodynamic measure tied directly to material transformation.

The action integral has been calculted from Current waveform using formula,

$g=\frac{1}{A^2}\int I(t)^2 dt$, where A is Initial cross-sectional Area of the foil and I is foil current. (1)

For a foil of length L and cross-sectional area A, if a current I (kA) is flowing through foil, then power delivered to the foil

$P(t) = (I(t))^2.\ \rho.\ L/A$, where ρ is electrical resistivity of foil. (2)

If u is the specific internal energy of foil and m is the mass of foil then total internal energy

$U = u.\ m$ (3)

Since power delivered to the foil $P(t) = dU/dt$, from (2) & (3) it can be written as

$(I(t))^2.\ \rho\ .\ L/A = m.\ du/dt$ (4)

$m = \sigma\ .A.L$ , $\sigma$ - mass density of metal foil.

Hence,

$\frac{1}{A^2}\int (I(t))^2.\ dt = \rho/\ \sigma\ .\int de$ = integral of specific action (5)

Under near-adiabatic conditions, the deposited Joule energy (LHS) is therefore purely a material dependent property (RHS). Therefore, particular values of action correspond to characteristic thermodynamic states such as, onset of melting, complete melting, onset of vaporization and explosive vapor burst/plasma formation.

Relating the specific action integral to material phase transitions is often done through a widely used Tucker-Toth Model, in exploding wire/foil studies. The comparison with the Tucker–Toth model provides a quantitative validation that the observed waveform features genuinely correspond to physical phase transitions in the exploding foil. The important point is that the Tucker–Toth model predicts these transition points from material properties alone — density, resistivity, latent heats, heat capacity, etc. Hence, agreement between experiment and theory indicates that the waveform interpretation, carried out earlier is physically meaningful.

**Table 3. Comparison of Action integral values obtained experimentally and as obtained from tucker-toth model**

| Phase | Tucker–Toth | Experiment | Deviation |
|---|---|---|---|
| Melt beginning | 25238 | 31215 | 23.6% |
| Melting end | 32035 | 41547 | 29.6% |
| Vapor beginning | 48561 | 52523 | 8.1% |
| Burst/vaporization | 65776 | 56700 | 13.7% |

The action integral of transition points predicted from Tucket Toth model and estimated from our experimental current wave form are compared in Table-3. Agreement is reasonably good considering that exploding foil phenomena is a highly dynamic process involving rapidly changing geometry, nonuniform current distribution, thermal gradients and hydrodynamic expansion. If conditions are appropriate, the later stages see arc formation, partial ionization. The entire phenomena witness time-dependent resistivity. Given all these complexities, deviations of roughly 10–30% are actually fairly acceptable and are commonly encountered in exploding conductor experiments (24).

Melting stages show larger deviations. The beginning and end of melting deviate by ~24–30%, larger than the vapor onset value. The Tucker–Toth model assumes a comparatively idealized and spatially uniform conductor. Real exploding foils violate this assumption strongly during melting. Melting in an exploding foil is not a sharp equilibrium transition. Different regions melt at different times: edge regions heat differently from the centre, non-uniform current density produces localized hotspots, solid and liquid phases coexist and expansion begins before melting is complete. Another important factor is that aluminium resistivity changes strongly near melting, making the inferred transition point from the electrical waveform somewhat ambiguous. Hence the experimentally inferred action integral tends to differ more during this stage.

The agreement is closer near vapor formation (8.1% deviation). Vaporization produces a much sharper electrical signature in the waveform than melting. The reason is that vaporization destroys metallic continuity far more drastically than melting does. Consequently, the electrical response becomes easier to identify experimentally, leading to closer agreement with the model. Vaporization is the most important precursor to plasma formation and current restrike.

For the final burst/vaporization stage, the experimental action is lower than the Tucker–Toth prediction. This deviation indicates onset of plasma effects. The Tucker–Toth approach is fundamentally a thermal phase-transition model. However, once vapor ionization, electron avalanche, and arc channels, begin to appear, the system no longer behaves as a neutral vapor. In practice, plasma conductivity can develop before complete equilibrium vaporization is achieved. This can trigger restrike and arc formation earlier than the idealized model predicts, reducing the experimentally observed action required for breakdown. Thus, the lower experimental burst action actually reflects the transition from thermal material physics to plasma discharge physics.

**4.3 Transition to Plasma** The most striking and potentially novel observation in the present experiments is the reproducible (See Fig 4) transient structure observed in region 6, immediately preceding current restrike. This region is characterized by a sharp voltage dip followed almost immediately by an equally sharp reversal into a rapid voltage rise accompanied by a simultaneous increase in current (See Fig 5). Significantly, this feature disappears entirely in cases where the deposited capacitor-bank energy is insufficient to produce restrike and sustained plasma formation (Fig 6 and Fig 7). This strong correlation suggests that Region 6 is a direct electrical signature of the transition from exploding metal vapor to a dynamic electrical breakdown process involving transient arc formation.

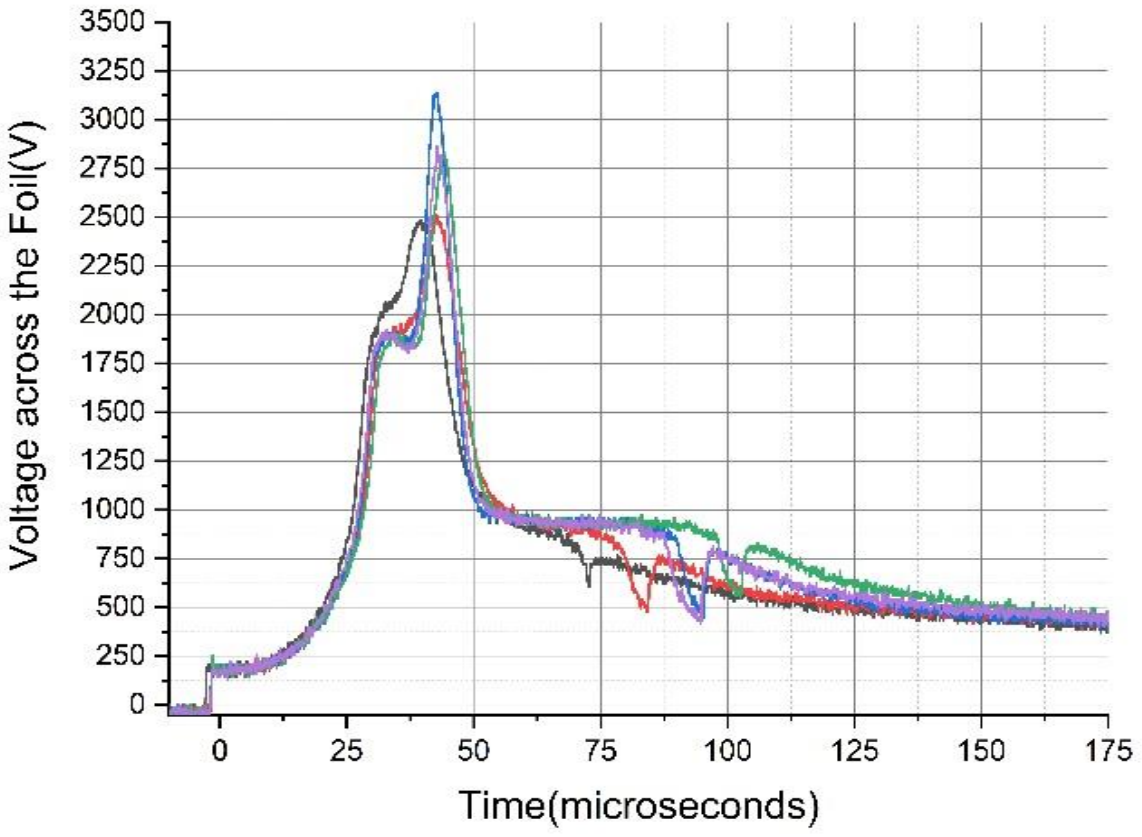

Figure 4.Novel dip signature reproducibly obtained in each experiment.

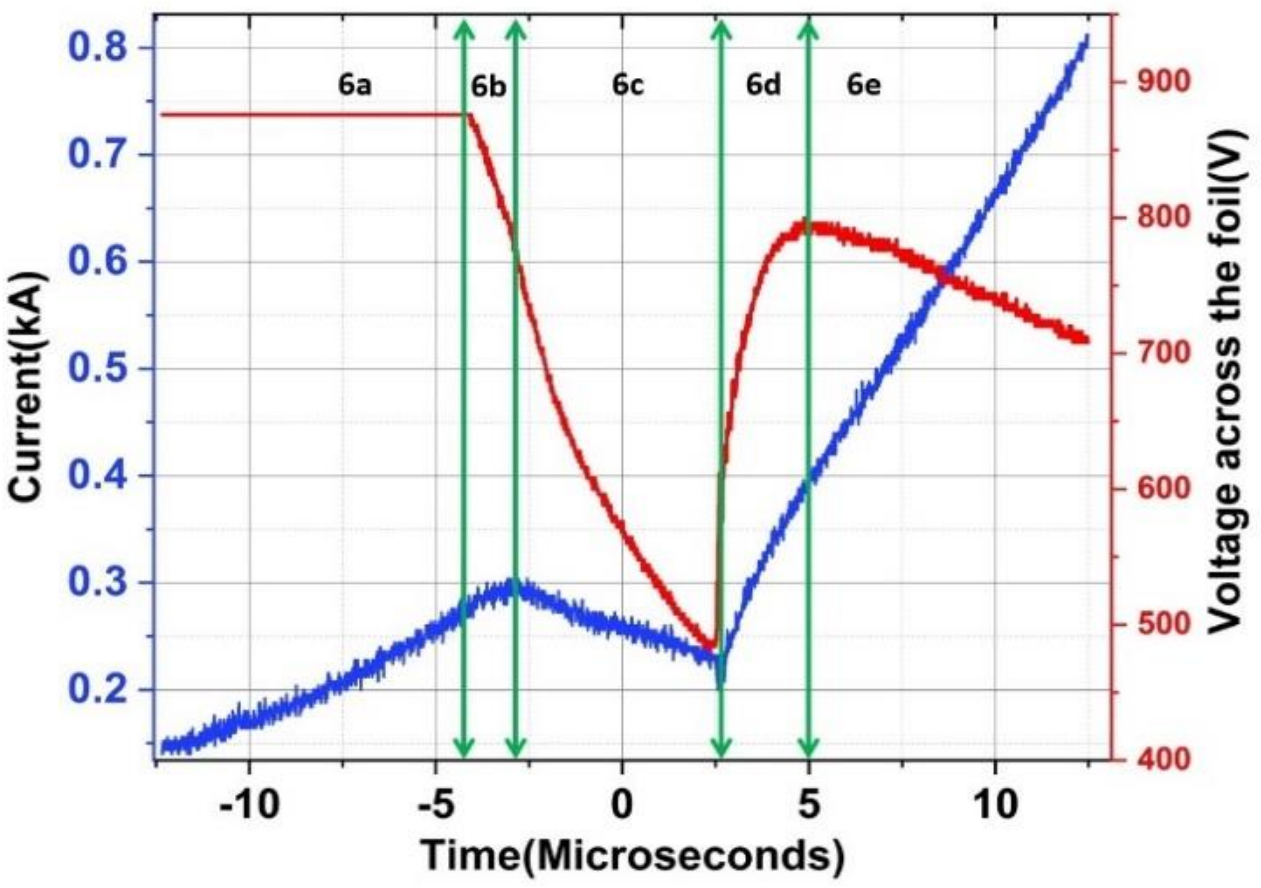

**Figure 5.Load Resistance in Region-6**

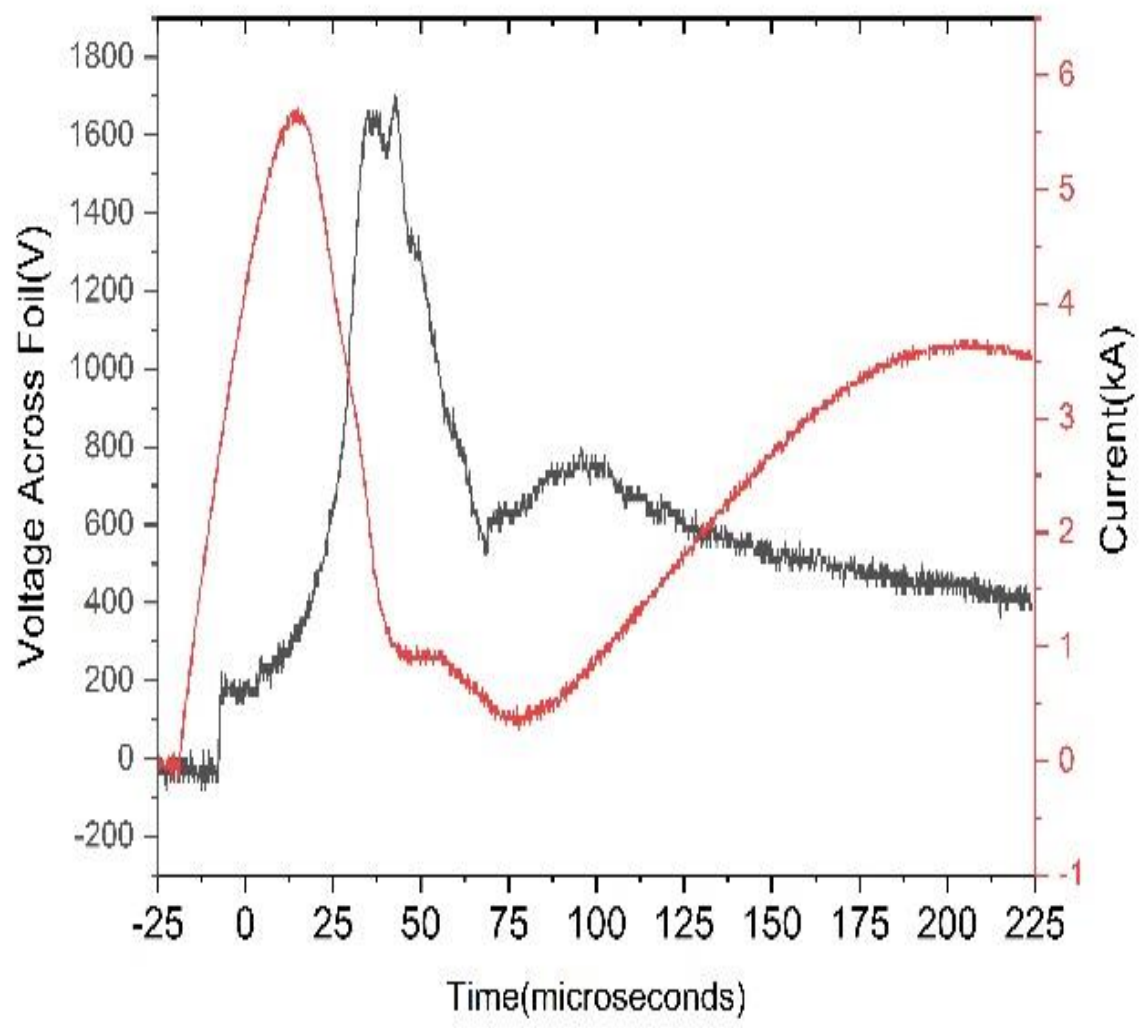

**Figure 6.Current and voltage when Al-foil exploded at 570J. Dip Signature in voltage across foil accompanied by a restrike of current**

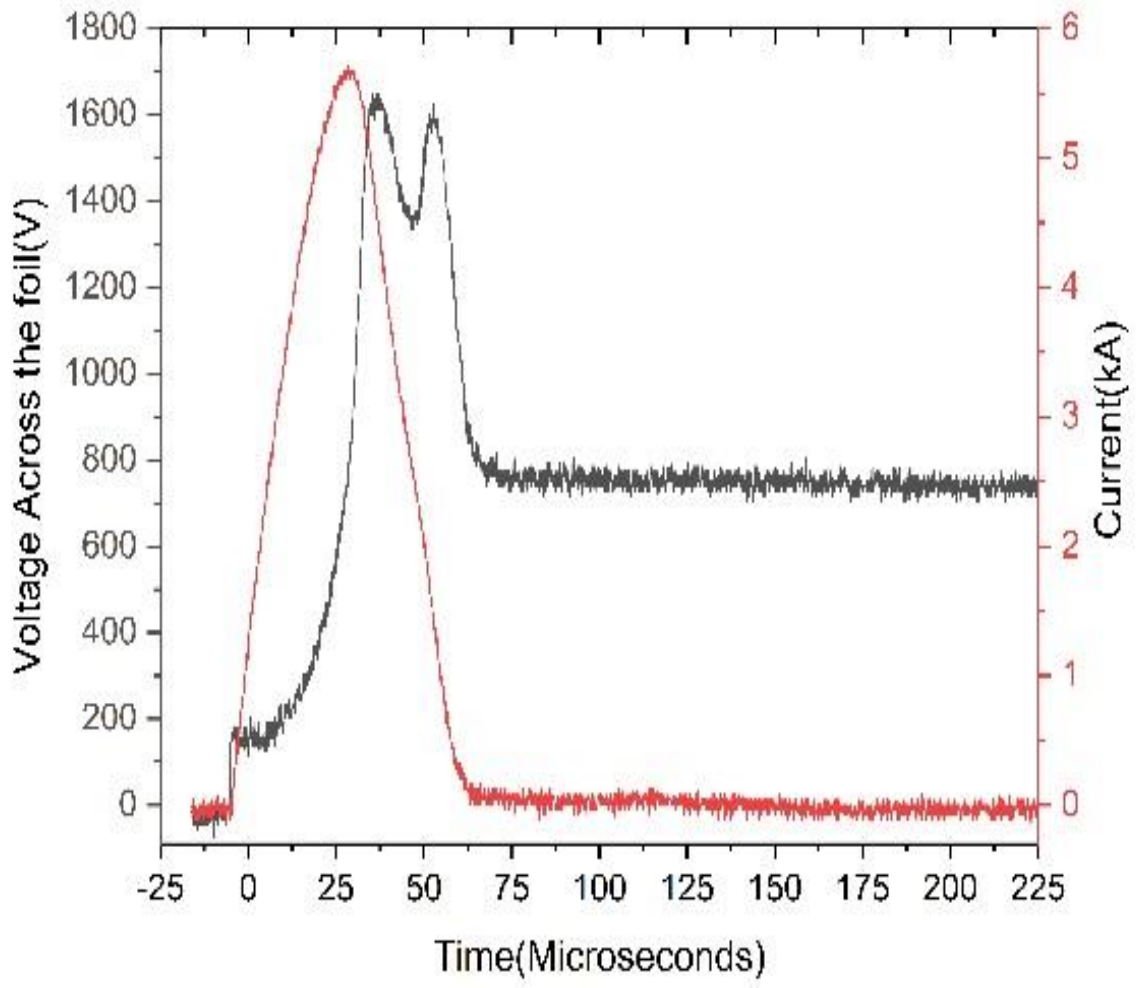

**Figure 7.Current and voltage when 42mg when exploded at 421 J. No Dip signature in voltage across foil and restrike of current.**

At the end of region 5, the metallic continuity of the foil has already been destroyed due to vaporization. The current is then forced to flow through a highly non-uniform mixture of dense metal vapor, liquid droplets, partially ionized species, and expanding low-density regions. Under these conditions, strong localized electric fields develop within the vapor column with the onset of localized electrical breakdown and filament formation within the metal vapor. Once a small region becomes weakly ionized, its conductivity increases locally, causing current to preferentially channel through that region. This produces additional Joule heating and further ionization, leading to rapid filament growth. The sudden formation of a conductive plasma filament causes a sharp reduction in effective discharge impedance, resulting in the observed voltage dip. The extremely rapid timescale of the voltage collapse is due to the avalanche-like behaviour characteristic of electrical breakdown phenomena. Since the external circuit driving the pulsed power possesses finite inductance, the current cannot respond instantaneously, explaining why the voltage change is considerably sharper than the accompanying current variation.

However, the subsequent simultaneous rise in both voltage and current is particularly important because it cannot be explained using simple resistive arguments alone. If the plasma channel merely became more conductive, increasing current would normally be accompanied by decreasing voltage. The observed behaviour instead indicates that the discharge has transitioned into a dynamic plasma state where the measured voltage is governed not entirely by resistance.

The abrupt reversal in the voltage waveform suggests that the discharge undergoes a nonlinear dynamical transition rather than a continuation of the preceding thermally dominated evolution. The exceptionally short timescale of the transition, together with its reproducible correlation with current restrike, is consistent with the onset of a threshold instability involving rapid ionization and plasma-channel formation Whether this transition arises primarily from the pulsed nature of the driving circuit, the unique evolution of dense metal vapor into plasma, or a coupled plasma-circuit instability remains an open question. Addressing this issue will require time-resolved plasma diagnostics together with self-consistent multi-physics models incorporating phase change, ionization kinetics, plasma transport, and the external pulsed-power circuit.

**5. Discussion and Conclusions:** The electrical explosion of Al foils was investigated under two energy-deposition conditions to examine the transformation of the conductor from solid metal to plasma. At the higher input energy of approximately 1.9 kJ, the foil undergoes rapid heating and explosive destruction, and the successive processes of melting, vaporization and ionization are compressed into a short time interval. The dominant electrical signature is a large voltage spike immediately preceding current restrike, correlated with the explosive loss of metallic continuity and subsequent formation of a luminous conducting plasma. The high-energy experiment thus reveals the overall conductor-to-plasma transition but does not resolve the intermediate processes leading to restrike.

To probe these dynamics, the energy deposition was reduced to approximately 0.57 kJ while maintaining comparable circuit inductance, capacitance and discharge timescales. The reduced rate of energy deposition spreads the evolution over approximately (100 μs), allowing characteristic electrical signatures associated with phase changes to be identified. Comparison with Tucker–Toth predictions strengthens our interpretation that the observed waveform regions genuinely correspond to solid heating, melting, liquid heating, vaporization, plasma formation and restrike. Without such agreement, assigning the waveform features to specific phase transitions would have remained much more speculative. Overall, our data suggest that the experiment operates in a regime where phase transitions are sufficiently resolved electrically; while the Tucker–Toth model captures the dominant thermal physics reasonably well, deviations arise from real exploding-foil effects such as nonuniform heating, expansion dynamics, and early plasma formation.

A particularly significant observation is the transient electrical structure immediately preceding current restrike. The voltage undergoes a pronounced decrease followed by an exceptionally sharp reversal and increase, accompanied by a rising current and eventual restrike. This feature is qualitatively distinct from the comparatively gradual thermal and phase-transition signatures observed earlier in the waveform and is reproducibly associated with successful plasma formation. The temporal behaviour of those earlier stages, involving rapid thermodynamic changes, remains comparatively gradual because thermal diffusion, latent heat absorption, and phase coexistence tend to smooth the transitions. The abruptness of the transition preceding restrike however indicates that the system is no longer governed primarily by thermal evolution. Possible contributions include rapid ionization, arc or filament development, current-path reorganization, evolving plasma impedance, and nonlinear coupling between the discharge and the pulsed-power circuit. At present, these mechanisms remain conjectural.

The results therefore identify a narrow but potentially important dynamical interval between explosive loss of metallic conduction and establishment of plasma conduction. The high-energy experiments reveal this interval as a compressed voltage spike preceding restrike, whereas reducing the energy-deposition rate provides a means of resolving the underlying sequence of phase transformations and pre-restrike dynamics. This approach offers a useful basis for further experimental diagnostics and coupled material–plasma–circuit modelling aimed at understanding the fundamental processes governing plasma formation during electrical explosions.